# The 5D-4D correspondence for the magnetic field

MARIA CRISTINA NEACSU
Department of Physics
Technical University "Gh.Asachi"
Blv. Copou No. 22, Iasi - 6600
Romania
email: mneacsu@mt.tuiasi.ro

*Abstract:* 5D theory is an alternative model for understanding gravitational and electromagnetic interactions together. In this work we used the correspondence between 5D Einstein field equations with cosmological constant and the 4D Einstein equations with sources. We started with the principal fibre bundle $P(M/\Phi, U(1))$ metric and studied the case when the gauge potential $A_a$ corresponds to the magnetic field: $A_a = (0, 0, A_3, 0)$. We identified the base-space with the space-time of two models of universe: the Robertson-Walker - like model and the Schwarzchild - like model. With this ansatz, we followed an algoritm that permits to express the magnetic field in terms of the gauge potential $A_a$ and the scalar field $\Phi$. This algorithm seems to work very well if the scalar field is time dependent. When we analyzed its dependence on other coordinates, new terms it comes out on the 5D field equations. We obtained the translation between the effective 4D electromagnetic potential and the 5D gauge and scalar fields.

## 1. Introduction

The magnetic field is present almost any where in Universe and this suggests it could be of fundamental origin. This means that the magnetic field is a consequence of a more general scalar-gravito-electro-magnetic field [1].

The primordial cosmological large-scale magnetic fields were essential for the evolution of the very early universe. For example, a primordial magnetic field could provide a seed field for dynamo amplifications in disc galaxies and an aditional form of relativistic energy density at the epoch of cosmological nucleosynthesis. It also would introduce anisotropies into the expansion dynamics and increases the expansion rate of the universe [2].

Although large-scale magnetic fields would not survive an epoch of inflation, it could be generated and becomes again important at the end of inflation and the beginning of the gravitational collapse.

The magnetic field is also an essential constituent of our days Universe, since planets, like earth, stars and galaxies posses a magnetic field. A great amount of astrophysical objects in Cosmos, like pulsars, quasars or black holes, are gravitational bodies endowed with magnetic field and there is not yet a convincing explanation for it [3].

Einstein theory is a good model for describing gravitational interactions in the universe. When gravitation is interacting with electromagnetism, we have to use the Einstein-Maxwell theory. This not a unification theory, but rather a superposition one, where the electromagnetic field appears like a model and there is no explanation for its existence [4],[5].

All modern field theories of unification contain scalar fields. In the Superstring (SS) and the Kaluza-Klein (KK) theories, scalar fields are fundamental fields and the electromagnetism follows from the decomposition and dimensional reduction [6]. The five-dimensional (5D) approach to gravity is a natural way of understanding gravitational and electromagnetic interactions together. It is well known that the 5D solutions of Einstein field equations in vacuum, when projected into four-dimensional space-time correspond to solutions of Einstein equations with an energy-momentum tensor of a gravitational field coupled with an electromagnetic field [7].

The actual version of 5D KK theores is based on the mathematical structure of the orincipal fibre bundle. In this work we start with the metric of the principal fibre bundle $P(M/\Phi, U(1))$ and study the case when the electromagnetic potential cuadri-vector has one nonvanishing component $A_3$, which describe the magnetic field. We identified the base-space $M$ with the space-time of two of the most used models of universe. The first one is the spatially homogenous and non-rotating Robertson Walker model that is very good for describing the evolution and the nature of the primordial magnetic field. The second one is the Schwarzschild-like model, which characterizes the geometry of spherical astrophysical objects, endowed with significant magnetic fields

We will use the correspondence between 5D Einstein field equations with cosmological constant and the 4D Einstein equations with sources. We will obtain an effective 4D energy-momentum tensor using only the 4D part of the 5D Einstein equations with cosmological constant. We consider as source of the gravitation a perfect magneto-fluid and from this decomposition yields a natural expression for the magnetic field in terms of the gauge potentials e and the scalar field.

## 2. The geometry

High-dimensional relativity seems to be an elegant way for unifying all interactions in physics and the first idea has been more and more transformed from the original suggestion of KK. Wesson and Leon [7] showed by algebraic means that the 5D Kaluza-Klein equations without sources may be reduced to the Einstein equations with sources and proved that the extra part of the 5D geometry could be used appropriately to define an effective 4D energy-momentum tensor. Hongya and Wesson [8] shown that 5D black hole solutions of KK theory can be reduced to new exact solutions of 4D general relativity in which matter is a spherically symmetric anisotropic fluid of radiation. Patel and Naresh [9] obtained from the five-dimensional cylindrically symmetric space-time, by dimensional reduction, the radiation Friedman- Robertson -Walker (FRW) flat model. Matos [10] presented a method for generating exact solutions of Einstein field equations as harmonic maps using the Chiral formalism in 5D. These solutions represent exterior fields of a gravitational body with arbitrary electromagnetic fields and whose gravitational potential posses a Schwarzschild - like behaviour [11]. He further elaborated a model [1] for the magnetic dipole of static bodies based on a 5D gauge theory that in 4D corresponds to a massive magnetic dipole coupled to a massless scalar field. Montesino and Matos [12] demonstrated that in 5D KK theories, when the lagrangian density reads: $\mathcal{L}_5 = R_5 + \Lambda$, with $\Lambda$ a 5D cosmological constant, after dimensional reductions and conformal transformations, one obtain: $\mathcal{L}_4 = -R_4 + 2(\nabla\Phi)^2 + e^{-2\alpha\phi}\Lambda$, where $\Phi$ is the scalar field, which introduce an exponential potential, and the coupling constant $\alpha$ has the value and $\alpha^2 = 3$.

In this work the attention is focused upon the five-dimensional relativity, which unifies gravitation with electromagnetism. The version adopted here considers that the 5D Riemannian space $P$ is a principal fiber bundle with typical fiber $S^1$ the circle. This version is more convenient for three reasons: it is a natural generalization of $U(1)$ - gauge theory (Maxwell theory) to curved space-times, there is no necessity to recur to the so-called Kaluza-Klein or to the n-mode ansatz and it is no need to impose any restrictions to the functional dependence of the metric terms on $P$. The geometry used is shortly explain bellow.

The formalism of gauge theory [13] is described in the framework of fibre bundle, which is a collection of elements $(P, \pi, M, F)$, explicitly given by three differentiable manifolds: the principal bundle $P$, the base manifold $M$ (usually taken to be the space-time manifold with metric $g$), the typical fibre $F$ which embodies the gauge freedom being a structure Lie group with transition functions that acts on $F$ from the left and the projection $\pi: P \to M$ whose inverse image $\pi^{-1}(p) \equiv F_p$ is the fibre at $p$. A certain gauge correspond to a certain section of $P$ and the gauge transformations are vertical automorphism $f_p : F_p \to F_p$ that change the section according to $s: f_p \circ s$. The gauge fields potentials are given by connection forms $\omega$ on $P$ that specifies the way in which a point of $P$ is parallel transported along a curve lying in the base manifold $M$ and yields corresponding curvatures or field strengths $\Omega$. In practice one uses their section-dependent counterparts $s*\omega$ and $s*\Omega$ defined on the base manifold. The matter fields are forms on the base manifold with values in a vector space $U$ of an associated vector bundle and the elements of the corresponding frame bundle constitute the reference frames used to decompose the matter fields with respect to $U$.

Let $U \subset M$ be an open subset of $M$. If $\phi: F \to U \times F$ is a trivialisation of an open subset of $F$ then the physical quantities can be mapped into the set $U \times F$ through the trivialisation $\phi$, which means that the total space $P$ looks locally like the direct product of $M$ and $F$. Let be:

$$\tilde{g} = \Phi^2 \hat{\omega} \otimes \hat{\omega} \qquad (1)$$

a metric on $F$ and:

$$\hat{g} = \eta_{AB} \hat{\omega}^A \otimes \hat{\omega}^B = \eta_{ab} \hat{\omega}^a \otimes \hat{\omega}^b + \Phi^2 \hat{\omega} \otimes \hat{\omega} \qquad (2)$$
$$A, B = 1,..5; a, b = 1,..4$$

the metric on $P=H \otimes V$ where $V$ is the vertical space and $H$ its complement. Let $\{e_a\}$ be the projection of the complement base $\{\hat{e}_a\}$ such that $d\pi(e_a)=e_a$ and $\pi_1$ be the first projection: $\pi_1: U \times F \to U$, $\pi_1(x,y)=x$. Thus, because of the identity $\pi = \pi_1 \circ \phi$, one finds that the corresponding base in the trivialisation set $U \times F$ is:

$$d\phi(\{\hat{e}_a, e\}) = \{e_a - A_a(\partial/\partial y), (\partial/\partial y)\} \qquad (3)$$

whose dual base is $\{\omega^a, dy+A_a\omega^a\}$. Thus, the corresponding metric on $U \times F$ will be:

$$\check{g} = \eta_{ab} \hat{\omega}^a \otimes \hat{\omega}^b + \Phi^2 (A_a\omega^a + dy) \otimes (A_b\omega^{ba} + dy) \qquad (4)$$

where it can be seen that $s*(\omega)=A_a\omega^a$, which means that $A_a\omega^a$ are the pullback components of the one-form connection $A$ through a cross section $s$. Since the group $U(1) \cong F$ is acting on $P$, there exist an isometry $Is:P \to P$, $(x^a,y)=(x^a,y+2\pi)$ such that $Is*\check{g}=\check{g}$. This implies the existence of a Killing-vector $\kappa$ and therefore we can be choose a coordinate system where the metric components of $\check{g}$ do not depend on $x^5=y$. In the gauge-theory, the action of $U(1)$ on $P$ means that there are electromagnetic interactions on $P$, which implies that there is a coordinate system where the metric components do not depend $x^5$. Then we could eliminate the dependence on coordinate $x^5=y$ of all the quantities involved in our study.

## 3. The field equations

We consider the scalar field coupled with gravity that yields from the KK theory. This means that we have to normalize $g_{ab} \to g_{ab}/\Phi$, the normalized potentials being now the physical gravitational potentials.

The scalar field $\Phi$ corresponds to the radius and $y$ to the coordinate of the internal space $U(1)$. It is not clear how to interpret physically any higher dimension. Interpretation of the fifth dimension has been done as a "massless scalar field" which can be or not associated with a fluid density [7], or as a "magnetic mass" [1]. Also interpretation has been done as a "fifth geometric property" which shows up near horizon. The scalar field was interpreted as "dialaton" or "inflation" [6] field and also was connected to the "dark matter" [14].

Our intention is to reduce the 5D field equations for vacuum with cosmological constant on the principal fibre bundle $P$, to 4D Einstein equations on $M$ without cosmological constant and with perfect charged fluid as source [15], [16]. Montesino and Matos [12] obtained the 4D energy-momentum tensor of a perfect fluid in terms of $\Phi$, using the following identification between the velocity 4-vector and the scalar field:

$$u_a \equiv \frac{\Phi_{;a}}{\sqrt{-\Phi_{;a}\Phi^{;a}}} \qquad (5)$$

If we choose a preferred vector field $u_a$ corresponding to a congruence of wordlines known as the "fundamental fluid-flow lines", that carry a family of special observers, namely the "fundamental observers", then $u_a$ is a timelike normalized ($u_au^a=-1$) vector field [17]. In this case, a dependence on the time coordinate of the scalar field will be imposed: $\Phi=\Phi(t)$, which means that the scalar field includes the time evolution of the Universe.

### 3.1. 5D-4D RW-like spacetime with electromagnetic potential

For studying the cosmological magnetic field, the appropriate chose of the line element for of the base spacetime will be the RW-like one [18]:

$$ds^2 = a^2\left[dr^2 + r^2\left(d\theta^2 + \sin^2\theta\, d\varphi^2\right)\right] - dt^2 \quad (6)$$

with the electromagnetic gauge potentials:

$$A_a = (0,0,A_3,0) \quad (7)$$

After we normalize the metric potentials, the 5D metric (4) yields:

$$d\tilde{s}^2 = \frac{a^2}{\Phi}\left(dr^2 + r^2 d\theta^2\right) - \frac{dt^2}{\Phi} + \Phi^2 dy^2 + \left(\frac{a^2 r^2 \sin^2\theta}{\Phi} + \Phi^2 A_3^2\right)d\varphi^2 + 2\Phi^2 A_3 d\varphi dy \quad (8)$$

The problem is how to choose the dependence of the scalar and potential field on the metric coordinates. The metric (5) admits a Killing vector associated with the $\varphi$ coordonate, so we can eliminate the dependence on $\varphi$. To simplify the calculations, we can assume that all the quantities involved will depend only on $(r,t)$. If the gauge potential will depend on time, we will obtain the electric component for the electromagnetic field: $E_a = 2u^b A_{[b,a]}$. So, from the definition of the magnetic field:

$$B^a = \eta^{abcd} u_b A_{[d,c]} \quad (9)$$

we conclude that the proper dependence for the gauge potential will be: $A_3 = A_3(r,\theta)$.

We calculated the 5D Einstein field equations for empty space with cosmological constant:

$$G_{AB} = \Lambda g_{AB}, A,B = 1,..5 \quad (10)$$

and the 4D Einstein-Maxwell field equations with a magneto-fluid as source:

$$G_{ab} = 8\pi T_{ab}, T^b_{a;b} = 0$$
$$F_{[ab;c]} = 0, F^b_{a;b} = J_a \quad (11)$$

within the GRTensor package [19]. The total energy-momentum tensor is: $T_{ab} = T^f_{ab} + T^{em}_{ab}$, where

$$T^f_{ab} = (\rho + p)u_a u_b + p g_{ab} \quad (12)$$

is the perfect fluid tensor, $p$ is the pressure, $\rho$ is the energy density, and:

$$T^{em}_{ab} = \frac{1}{4\pi}\left(F_{ac}F^c_b + \frac{1}{4}g_{ab}F^{cd}F_{cd}\right) \quad (13)$$

is the electromagnetic field tensor, $F_{ab} = 2A_{[b,a]}$ being the Maxwell tensor.

We studied the following cases (comma means derivation with $t$, dot is for derivation with $r$, and star is for derivation with $\theta$):

#### 3.1.1. a, Φ, A₃, p, ρ=f(t)

In this case the magnetic field is vanishing and we have only the electric part of the electromagnetic field. The 5D-4D reduction is possible because the symmetry of signs is respected and we find the expression of the electric field in terms of the scalar field:

$$A'^2 \to \frac{\Phi^3 A_3'^2}{4} \quad (14)$$

From the conservation of the electromagnetic tensor we have the restriction:

$$\frac{a'}{a} + \frac{A_3''}{A_3'} = 0 \quad (15)$$

A similar, but not identical relation is obtained from the field equation:

$$G^3_5 = \Lambda g_{AB} \Rightarrow 3\frac{\Phi'}{\Phi} + \frac{a'}{a} + \frac{A_3''}{A_3'} = 0 \quad (16)$$

and we conclude that the scalar field contributes with a supplementary term.

#### 3.1.2. a, Φ, A₃, p, ρ=f(r,t)

In this case we have both magnetic and electric field, and from the identification we find:

$$A'^2 \to \frac{\Phi^3 A_3'^2}{4} \text{ and } \dot{A}^2 \to \frac{\Phi^3 \dot{A}_3^2}{4} \quad (17)$$

The conservation of the electromagnetic tensor yields:

$$a^2 \frac{A_3'}{A_3}\left(\frac{a'}{a} + \frac{A_3''}{A_3'}\right) + \frac{A_3^*}{A_3}\left(\frac{a'}{a} - \frac{\ddot{A}_3}{\dot{A}_3}\right) = 0 \quad (18)$$

and the $G^3_5$ component gives:

$$a^2 \frac{A_3'}{A_3}\left(\frac{a'}{a} + \frac{A_3''}{A_3'} + 3\frac{\Phi'}{\Phi}\right) + \frac{A_3^*}{A_3}\left(\frac{a'}{a} - \frac{\ddot{A}_3}{\dot{A}_3} - 3\frac{\dot{\Phi}}{\Phi}\right) = 0 \quad (19)$$

From the correspondence of the 5D-4D nondiagonal terms of the field equations, yields:

$$A_3' \dot{A}_3 \to \frac{\Phi^3 A_3' \dot{A}_3}{4} - \frac{3}{4}\frac{\Phi'}{\Phi}\frac{\dot{\Phi}}{\Phi} r^2 a^2 \sin^2\theta \quad (20)$$

### 3.1.3. a, Φ, p, ρ=f(t), A₃ =f(r,θ)

This is the case only with magnetic field and the model seems to work better, because from the electromagnetic tensor conservation and the $G^3{}_5$ component, we obtain the same relation:

$$\ddot{A}_3 r^2 + A_3^{**} = A_3^* \cot\theta \qquad (21)$$

the sign symmetry is conserved, and from the dimensional reduction yields:

$$\dot{A}^2 \to \frac{\Phi^3 \dot{A}_3^2}{4} \text{ and } A^{*2} = \frac{\Phi^3 A_3^{*2}}{4} \qquad (22)$$

The nondiagonal tems provide the translation:

$$\dot{A}_3 A_3^* \to \frac{\Phi^3 \dot{A}_3 A_3^*}{4} \qquad (23)$$

### 3.2. 5D-4D Schwarzchild-like spacetime with electromagnetic potential

The exterior of a spherical, magnetized, astrophysical object can be modeled with the static, spherically symmetric Schwarzchild-like metric [5]:

$$ds^2 = \frac{dr^2}{a} + r^2(d\theta^2 + \sin^2\theta d\varphi^2) - adt^2 \qquad (24)$$

with the same form for the electromagnetic gauge potential as in equation (6). This metric has a higher symmetry than the RW one and admits two Killing vectors associated with the (φ, t) coordinates, so we can impose for the physical quantities only the dependence on (r,θ).

For this choose of the base space-time line element, the 5D metric (4) becomes:

$$d\tilde{s}^2 = \frac{dr^2}{a\Phi} + \frac{r^2 d\theta^2}{\Phi} - \frac{adt^2}{\Phi} + \Phi^2 dy^2 + \left(\frac{r^2\sin^2\theta}{\Phi} + \Phi^2 A_3^2\right)d\varphi^2 + 2\Phi^2 A_3 d\varphi dy \qquad (25)$$

The problem is that our metric is static, but the scalar field must depend on time, so we will be forced to consider the time dependence for the other quantities involved, too.

We have to calculated the same equations: $G_{AB}=\Lambda g_{AB}$ for 5D and the 4D Einstein-Maxwell field equations with a magneto-fluid as source: $G_{ab}=8\pi T_{ab}$. The cases that we have studied are:

### 3.2.1. a, A₃, p, ρ=f(r), Φ=f(t)

From the electromagnetic tensor conservation and the $G^3{}_5$ component, we obtain the identity:

$$\frac{\dot{a}}{a} + \frac{\ddot{A}_3}{\dot{A}_3} = 0 \qquad (26)$$

and the scalar field normalizes in the same way the magnetic potential:

$$\dot{A}^2 \to \frac{\Phi^3 \dot{A}_3^2}{4} \qquad (27)$$

### 3.2.2. a, Φ, A₃, p, ρ=f(r,t)

The electromagnetic tensor conservation yields:

$$a^2 \frac{\dot{A}_3}{A_3}\left(\frac{\dot{a}}{a} + \frac{\ddot{A}_3}{\dot{A}_3}\right) + \frac{A_3'}{A_3}\left(\frac{a'}{a} - \frac{A_3''}{A_3'}\right) = 0 \qquad (28)$$

and from the $G^3{}_5$ component, we obtain:

$$a^2 \frac{\dot{A}_3}{A_3}\left(\frac{\dot{a}}{a} + \frac{\ddot{A}_3}{\dot{A}_3} + 3\frac{\dot{\Phi}}{\Phi}\right) + \frac{A_3'}{A_3}\left(\frac{a'}{a} - \frac{A_3''}{A_3'} - 3\frac{\Phi'}{\Phi}\right) = 0 \qquad (29)$$

The 5D-4D correspondence for the nondiagonal components gives:

$$\frac{A_3' \dot{A}_3}{ar^2 \sin^2\theta} \to \frac{\Phi^3 A_3' \dot{A}_3}{4ar^2\sin^2\theta} - \frac{3}{4}\frac{\Phi'}{\Phi}\frac{\dot{\Phi}}{\Phi} \qquad (30)$$

### 3.2.3. a, p, ρ=f(r), Φ=f(t), A₃ =f(r,θ)

This is again the case with a straightforward identification, and the electromagnetic tensor conservation is identical with the $G^3{}_5$ component vanishing condition:

$$r^2 a\ddot{A}_3 + r^2 \dot{a}\dot{A}_3 + A_3^{**} = A_3^* \cot\theta \qquad (31)$$

The magnetic potentials admit the same translation (21), and the nondiagonal components of the electromagnetic tensor have the following direct correspondence on the nondiagonal components of the 5D Einstein tensor:

$$\dot{A}_3 A_3^* \to \frac{\Phi^3 \dot{A}_3 A_3^*}{4} \qquad (32)$$

## 4. Conclusions

In this paper we had shown that the field equations with cosmological constant on the principal fibre bundle with nonvanishing magnetic potential, can lead to the energy-momentum tensor of a charged perfect fluid. This algorithm seems to work very well if the scalar field is time dependent. When we analyzed its dependence on other coordinates, new terms it comes out on the 5D field equations. If we totally eliminate the dependence on time of the scalar field, the sign symmetry is broken and the correspondence is not possible.

The study of the above problems involves a lot of interesting aspects besides the ones analyzed here. It can be pointed out the study of the $G_{55}$ component in order to understand the information that we receive from the fifth dimension. It is also convenient a deeper investigation on the relationship between this model and the string theory for low energies. In any case, it is worth to explore ways of interpreting the properties of the 5D solutions in a 4D word.